\documentclass[journal]{IEEEtran}

\ifCLASSINFOpdf
\else
   \usepackage[dvips]{graphicx}
\fi
\usepackage{url}

\usepackage{graphicx}
\usepackage[margin=1.5cm]{geometry}
\usepackage{graphicx,subcaption,lipsum}
\usepackage{float}
\usepackage{amsfonts}
\usepackage{amsmath}
\usepackage{color}
\usepackage{url}
\usepackage{framed}
\usepackage{pdfpages}
\usepackage{empheq}
\usepackage{amssymb}
\usepackage{bm}
\usepackage[normalem]{ulem}

\newtheorem{lemma}{Lemma}[section]
\newtheorem{proposition}{Proposition}[section]

\newtheorem{remark}{Remark}[section]

\newcommand{\E}{\mathbb{E}}

\usepackage{xcolor}

\begin{document}

\title{Quickest Change Detection in Continuous-Time in Presence of a Covert Adversary}

\author{Amir Reza Ramtin, Philippe Nain, Don Towsley, \IEEEmembership{Life Fellow, IEEE}
\thanks{This research is supported by the DEVCOM Army Research Laboratory under Cooperative Agreement W911NF-17-2-0196 (IoBT CRA) and the National Science Foundation under Grant ECCS-2148159.}
\thanks{The authors are grateful to Professor Venugopal V. Veeravalli for helpful comments and suggestions.}
\thanks{A. R. Ramtin and D. Towsley are with the University of Massachusetts at Amherst, MA, USA (email: aramtin@cs.umass.edu, towsley@cs.umass.edu).}
\thanks{P. Nain is with Inria, France (email: philippe.nain@inria.fr).}}

\maketitle

\begin{abstract}
We investigate the problem of covert quickest change detection in a continuous-time setting, where a Brownian motion experiences a drift change at an unknown time. Unlike classical formulations, we consider a covert adversary who adjusts the post-change drift $\mu = \mu(\gamma)$ as a function of the false alarm constraint parameter $\gamma$, with the goal of remaining undetected for as long as possible. Leveraging the exact expressions for the average detection delay (ADD) and average time to false alarm (AT2FA) known for the continuous-time CuSum procedure, we rigorously analyze how the asymptotic behavior of ADD evolves as $\mu(\gamma) \to 0$ with increasing $\gamma$. Our results reveal that classical detection delay characterizations no longer hold in this regime. We derive sharp asymptotic expressions for the ADD under various convergence rates of $\mu(\gamma)$, identify precise conditions for maintaining covertness, and characterize the total damage inflicted by the adversary. We show that the adversary achieves maximal damage when the drift scales as $\mu(\gamma) = \Theta(1/\sqrt{\gamma})$, marking a fundamental trade-off between stealth and impact in continuous-time detection systems.
\end{abstract}


\IEEEpeerreviewmaketitle

\section{Introduction}
\label{sec:intro}
Quickest change detection (QCD) has long been a fundamental statistical technique for rapidly identifying changes in the underlying statistical properties of observed processes, with numerous critical applications ranging from quality control and finance to security and environmental monitoring \cite{Poor, Shiryaev, Tartakovsky}. Traditionally formulated in discrete-time settings, QCD methods have also been extensively developed and analyzed within continuous-time frameworks, where processes are monitored continuously rather than at discrete intervals. Continuous-time detection is particularly advantageous in settings requiring immediate responses to subtle or rapid shifts, as it allows for instantaneous detection of changes, thereby reducing reaction times and potential damages \cite{Beibel,Poor,Shiryaev}.

A classical and widely used method in continuous-time QCD is the continuous-time Cumulative Sum (CuSum) procedure, initially studied by Lorden \cite{Lorden} and subsequently generalized by Moustakides \cite{Moustakides}. The continuous-time CuSum is known for its optimality under Lorden's minimax criterion, which aims to minimize the worst-case expected detection delay while controlling the average time between false alarms. Under this criterion, the CuSum procedure continuously accumulates evidence (in the form of log-likelihood ratios) and signals a change once this cumulative evidence crosses a predetermined threshold \cite{Moustakides}, \cite{Poor}. Precise characterizations of performance metrics, such as average time to false alarm (AT2FA) and average detection delay (ADD), are available in closed form for standard continuous-time models \cite{Moustakides}.

An important application of QCD is detecting  the presence of an adversary \cite{Xie2021}. However, classical analyses typically assume that the adversary is unaware of the detection mechanism. In adversarial environments, these assumptions may no longer hold. Specifically, an 
adversary, aware of the detection mechanism, may strategically adapt their behavior in response to detection thresholds. In this scenario, the adversary's statistical signature post-change increasingly resembles the pre-change condition as the false-alarm constraint (quantified by a parameter $\gamma$) becomes increasingly stringent. Such adaptive adversarial behavior challenges traditional assumptions and necessitates a deeper theoretical analysis of covert adversaries operating in a continuous-time setting.

The notion of covertness, first introduced by Bash et al. in the context of covert communication \cite{bash2013limits}, captures the ability of a transmitter to remain undetectable by an observing warden during communication. Most existing work assumes that the detector knows the exact start time of any potential transmission. In contrast, we relax this assumption and extend the notion of covertness to a sequential detection framework, where adversaries deliberately behave in a way that keeps their activity concealed over extended periods. From the perspective of sequential detection, this gives rise to a fundamentally different adversarial strategy. Unlike traditional adversaries whose actions lead to detectable statistical deviations---enabling timely detection---covert adversaries aim to remain hidden by ensuring that ADD scales with the same order as AT2FA, thereby rendering the detector ineffective \cite{huang_covert_2021}.

While prior work has studied covertness in discrete-time sequential detection \cite{ chang_covert_2021,huang_covert_2021, ramtin2024quickest}, the continuous-time counterpart of this problem remains relatively unexplored. 
The works in \cite{chang_covert_2021, huang_covert_2021} have been largely confined to covert wireless communication scenarios, with assumptions tailored to that domain---such as specific channel models or power constraints---and have primarily focused on communication-centric performance metrics. In contrast, \cite{ramtin2024quickest} considers an adversarial setting against the CuSum procedure under non-stationarity and without assuming that the AT2FA constraint is known to the adversary.

In this paper, we rigorously investigate covert adversaries in a continuous-time CuSum framework, where the post-change drift parameter is allowed to depend explicitly on the false-alarm constraint parameter $\gamma$. Specifically, we examine scenarios in which the drift parameter of the underlying continuous-time stochastic process approaches its pre-change value as $\gamma \to \infty$. Our analysis leverages classical exact results for continuous-time CuSum \cite{Moustakides}, extending their applicability to the asymptotic regime characterized by diminishing statistical differences between pre- and post-change processes. We derive expressions for ADD under various regimes, determined by the rate at which the post-change drift vanishes with increasing $\gamma$. Our results demonstrate that the classical formulas for the ADD of CuSum, which assume a fixed drift, no longer hold in this asymptotic regime and must be carefully revised to reflect the vanishing signal structure introduced by a covert adversary.

We characterize precise conditions under which an adversary can maintain covertness and derive the corresponding scaling laws. Within the Brownian motion framework, adversarial behavior manifests through changes in the drift parameter. We analyze how the scaling of this drift---relative to the run-length constraint---affects the ability to remain covert. We further quantify the adversarial damage, defined as the product of the drift magnitude and the average detection delay, and determine the regimes in which this damage is maximized while preserving covertness.

The remainder of the paper is structured as follows. In the next section, we briefly review the continuous-time CuSum procedure and Lorden's minimax detection framework. Section \ref{sec:covert} develops our primary theoretical results. We then provide numerical evaluations that substantiate our theoretical results. Finally, we conclude with a summary and discuss future directions.


\section{Review of quickest change detection in continuous-time}
\label{sec:review}
The material presented in this section can be found in \cite[Section 6.4]{Poor}. See also \cite{Beibel, Lorden, Moustakides, Shiryaev}.
In the continuous-time setting, the stochastic process $\{Z_s,s\geq 0\}$ is continuously observed with the following dynamics
\begin{equation}
\label{brownian-model}
dZ_s=\left\{\begin{array}{ll}
dW_s, &\mbox{$s\leq t$},\\
\mu ds +dW_s,&\mbox{$s>t$,}
\end{array}
\right.
\end{equation}
where $\{W_s, s\geq 0\}$ is a standard Brownian motion. The drift parameter $\mu$ is known and (without loss of generality) positive. Denote by  ${\cal F}_s=\sigma(Z_u, 0\leq u\leq s)$ the smallest $\sigma$-field  with respect to which process $\{Z_u,0\leq u\leq s\}$ is measurable.
In (\ref{brownian-model}),  $t$ is the (unknown) time when the change occurs. In words, before the change point the process $\{Z_s, s\geq 0\}$ is a Brownian motion and after the change point it is  a particular instance of an Ito process. No prior distribution on the change point $t$ is assumed (non-Bayesian setting).

The goal of quickest change detection is to minimize the delay between when a change occurs and when it is detected, while maintaining a low false alarm rate.  More specifically, a
change detection algorithm is a stopping time $T$ with respect to the filtration  $\{ {\cal F}_s, s\geq 0\}$,  that is $\{T\leq s\}\in {\cal F}_s$ for all $s\in [0,\infty)$. If $T\geq t$ 
a delayed detection is made otherwise a false alarm has occurred. 

We denote by ${\cal T}$ the set of all stopping times with respect to $\{ {\cal F}_s, s\geq 0\}$.  Throughout $\E_t$ $(t\geq 0$) is the expectation operator associated with the model in (\ref{brownian-model}); in particular, the operator $\E_0$ is associated with  the situation when a change occurs at time $t=0$ whereas the operator $\E_\infty$ is associated with the situation where no change occurs.

In this work we consider the following optimization problem, due to Lorden  \cite{Lorden}, to find the best stopping time $T$ in ${\cal T}$:
\begin{equation}
\label{opt-pb}
\inf_{T\in {\cal T}} J(T)\quad \hbox{subject to } \E_\infty[T]\geq \gamma,
\end{equation}
with $\gamma>0$, where 
\begin{equation}
\label{worst-case}
J(T)=\sup_{t\geq 0}\,\hbox{ess sup }\E_t[\max\{0,T-t\}\,|\,{\cal F}_t]
\end{equation}
characterizes the worst-case expected delay. In (\ref{worst-case}) ess sup stands for essential supremum. The constraint in (\ref{opt-pb}) expresses
the fact that the false alarm rate should not exceed $1/\gamma$. Denote by $n(\gamma)=\inf_{T\in {\cal F} : \E_\infty[T]\geq \gamma}d(T)$, the optimal worst-case average detection delay.

It is known  \cite[Section 6.4.1]{Poor} that the stopping time $T_h:=\inf\{ s\geq 0\,:\, Y_s\geq h\}$ with $Y_s:=U_s-M_s$
where $U_s:= \mu Z_s-\frac{1}{2}\mu^2 s$ and $M_s:=\inf_{0\leq r\leq s} U_r$,  with $h$ selected so that $\E_\infty[T_h]=\gamma$, solves (\ref{opt-pb}). 
It is also known that for any $T\in {\cal T}$, the worst-case detection delay occurs when $t=0$, namely, $J(T)=\E_0[T]$, which in turn implies from the optimality of $T_h$ that
\begin{equation}
\label{opt-solution}
n(\gamma)=\E_0[T_h].
\end{equation}
In addition \cite[Prop. 6.8, p. 146]{Poor},
\begin{align}
\E_\infty[T_h]&=\frac{2}{\mu^2}(e^h -h-1), \label{Einfinity}\\
\E_0[T_h]&=\frac{2}{\mu^2}(e^{-h} +h-1).\label{E0}
\end{align}
In \cite{Moustakides} Moustakides generalized the model in (\ref{brownian-model}) to the case when $\mu$ depends on $s$ and proposed a generalized version 
of Lorden's minimax criterion to find the optimal detection delay.

In the following, we will investigate the model in (\ref{brownian-model}) when $\mu$ depends on $\gamma$ and $\lim_\gamma \mu(\gamma)=0$.

\section{Conditions for covertness} 
\label{sec:covert}

In this section,  we conduct an asymptotic analysis of the model in (\ref{brownian-model})  as $\gamma\to\infty$, for the case where 
the drift $\mu$ depends on $\gamma$, denoted by $\mu(\gamma)$. We assume that $\mu(\gamma)>0$ for all $\gamma>0$ and 
$\lim_{\gamma\to\infty} \mu(\gamma)=0$. 

Notice that $n(\gamma)=\Theta(\gamma)$ can be interpreted as a covertness condition since it says that
$n(\gamma)$ is of the same order of magnitude as the largest admissible lower bound on the expected time between false alarms, given by $\gamma$. With this in mind, we address the following questions: (1) is it possible to have $n(\gamma)=\Theta(\gamma)$ and, if yes, (2) under what conditions? 


 
Our analysis will use the Lambert function. Recall that the Lambert function  is the solution of the equation $W(z)e^{W(z)}=z$  \cite{Corless}.  When $z$ is real, $W(z)e^{W(z)}=z$ has a solution if and only if $z\geq -e^{-1}$. When $z\geq 0$, this solution is
unique, given by the main branch  $W_0$ of $W$, and when $-e^{-1}\leq z<0$ there are two solutions given by the branches $W_0$ and $W_{-1}$ of $W$. In particular,
\begin{equation}
\label{branches-Lambert}
W_0(z)>-1 \quad \hbox{and}\quad W_{-1}(z)< -1\,\hbox{ for }-e^{-1}<z<0,
\end{equation}
and $W_0(-e^{-1})=W_{-1}(-e^{-1})=-1$. Last \cite[p. 350]{Corless}
\begin{equation}
\label{limit-Lambert}
W_{-1}(z)\sim \log(-z)-\log(-\log(-z)) \,\hbox{ as } z\to 0 \hbox{ with } z<0.
\end{equation}

\begin{proposition}[Asymptotic behavior of $n(\gamma)$]\hfill
\label{prop:asympt}
As $\gamma\to\infty$,
\begin{equation}
n(\gamma)\sim 
\left\{\begin{array}{ll}
\frac{2}{\mu(\gamma)^2}\log(\gamma \mu(\gamma)^2), &\mbox{if $\lim_{\gamma\to\infty} \gamma \mu(\gamma)^2=\infty$,}\\
 \frac{2}{\theta}G\left(\frac{1}{2}\theta \right)\gamma,&\mbox{if $\lim_{\gamma\to\infty}\gamma \mu(\gamma)^2 :=\theta$,}\\ 
 \gamma, &\mbox{if $\lim_{\gamma\to\infty}\gamma \mu(\gamma)^2 =0$,}
\end{array}
\right.
\label{limit-general-E0-infinity-zero}
\end{equation}
with $\theta\in (0,\infty)$, where 
\begin{equation}
\label{def-G}
G(x)=e^{1+x +W_{-1}(-e^{-1-x})}-W_{-1}(-e^{-1-x})-x-2, \quad x\geq 0
\end{equation}
\end{proposition}

The proof of Proposition \ref{prop:asympt} is based on Lemma \ref{lem:solving-constraint} stated below, whose proof is given in the appendix.

\begin{lemma}[Optimal threshold]\hfill
\label{lem:solving-constraint}

For every $\gamma>0$, the equation $\E_\infty[T_h]=\gamma$ has a unique solution in $[0,\infty)$, given by
\begin{equation}
\label{h-star}
h(\gamma)=-1-\frac{1}{2}\gamma\mu(\gamma)^2-W_{-1}\left(-e^{-1-\frac{1}{2}\gamma\mu(\gamma)^2}\right).
\end{equation}
\end{lemma}

{\em Proof of  Proposition \ref{prop:asympt}:}  From (\ref{opt-solution}), (\ref{E0}), (\ref{def-G}),  and (\ref{h-star}), we have
\begin{align}
n(\gamma)=&\frac{2}{\mu(\gamma)^2} G\left(\tfrac{1}{2}\gamma \mu(\gamma)^2\right).
\label{n-gamma}
\end{align}

Assume first that $\lim_{\gamma\to\infty} \gamma \mu(\gamma)^2 =\infty$. From (\ref{limit-Lambert}) we get 
$W_{-1}\left(-e^{-1-\frac{1}{2}\gamma\mu(\gamma)^2}\right)\sim \log(\gamma \mu(\gamma)^2)$ as $\gamma\to \infty$, yielding
$G\left(\tfrac{1}{2}\gamma \mu(\gamma)^2\right)\sim \log(\gamma \mu(\gamma)^2)$ as $\gamma\to\infty$, and subsequently,
\[
n(\gamma)\sim \frac{2}{\mu(\gamma)^2}\log(\gamma \mu(\gamma)^2), \quad \gamma\to\infty.
\]
If $\lim_{\gamma\to\infty} \gamma\mu(\gamma)^2 =\theta\in (0,\infty)$,  clearly from (\ref{n-gamma})
\begin{align*}
n(\gamma)\sim \frac{2}{\theta}G(\tfrac{1}{2}\theta)\gamma, \quad \gamma\to\infty.
\end{align*}
Assume now that $\lim_{\gamma\to\infty} \gamma \mu(\gamma)^2 =0$. Let us prove that $\lim_{x\downarrow 0}\frac{G(x)}{x}=1$, which will prove from
(\ref{n-gamma}) that $n(\gamma)\sim \gamma$ as $\gamma\to\infty$.

Since $G(0)=0$,  $\lim_{x\to 0\atop x>0}\frac{G(x)}{x}$ is the right hand derivative of $G(x)$ at $x=0$. For
$-e^{-1}<x<0$, $W_1(x)$ has a derivative, given by (Hint: differentiate $W_{-1}(x)e^{W_{-1}(x)}=x$)
 \begin{equation}
 \label{derivative-W-1}
W_1^\prime(x)=\frac{W_{-1}(x)}{x(1+W_{-1}(x))}.
 \end{equation}
Consequently,  for $x>0$, the derivative of $G(x)$ is given by
 \begin{align*}
G^\prime(x)=&e^{-1-x}W_{-1}^\prime(-e^{-1-x})\left(e^{1+x+W_{-1}(-e^{-1-x})} -1\right)\\
 &+e^{1+x+W_{-1}(-e^{-1-x})}-1,\\
=&  \left(-\frac{W_{-1}(-e^{-1-x})}{1+W_{-1}(-e^{-1-x})}+1\right) e^{1+x+W_{-1}(-e^{-1-x})}\\
&+ \frac{W_{-1}(-e^{-1-x})}{1+W_{-1}(-e^{-1-x})}-1,\\
=&\frac{e^{1+x+W_{-1}(-e^{-1-x})}-1}{1+W_{-1}(-e^{-1-x})} =-\frac{1}{W_{-1}(-e^{-1-x})}.
\end{align*}
Letting $x\to 0$ with $x>0$, we see that $1=\lim_{x\to 0\atop x>0}G^\prime(x)=\lim_{x\to 0\atop x>0}\frac{G(x)}{x}$, which concludes the proof
of Proposition  \ref{prop:asympt}.

According to our definition of covertness, given by $n(\gamma)=\Theta(n)$, Proposition \ref{prop:asympt} shows that an adversary is covert if $\mu(\gamma)\sqrt{\gamma}$ converges to a {\em finite} limit  as $\gamma\to\infty$.

\begin{remark}[Asymptotic behavior of optimal threshold $h$]
\begin{align*}
&\lim_{\gamma\to\infty}h(\gamma)=\\
&\left\{\begin{array}{ll}
\infty, &\mbox{if $\lim_{\gamma\to\infty}\gamma\mu(\gamma)^2 =\infty$},\\
-1-\frac{1}{2}\theta - W_{-1}(-e^{-1-\frac{1}{2}\theta}), &\mbox{if $\lim_{\gamma\to\infty}\gamma\mu(\gamma)^2 =\theta$},\\
1, &\mbox{if $\lim_{\gamma\to\infty}\gamma\mu(\gamma)^2 =0$}.
\end{array}
\right.
\end{align*}
The result when $\lim_{\gamma\to\infty}\gamma\mu(\gamma)^2 =\infty$ follows by applying the expansion in (\ref{limit-Lambert}) to $h(\gamma)$ in (\ref{h-star}), while
the result when $\lim_{\gamma\to\infty}\gamma\mu(\gamma)^2 =0$ follows from (\ref{h-star}) and $W_{-1}(-e^{-1})=-1$.
\end{remark}

\begin{remark}
If $\lim_{\gamma\to\infty} \mu(\gamma)=\mu_0\not=0$ then $n(\gamma)\sim \frac{2}{\mu)^2}\log \gamma$ by Proposition \ref{prop:asympt}.
In particular, this shows that  $n(\gamma)=\Theta(\log(\gamma))$ 
in continuous-time. 
\end{remark}

\begin{remark}
It is worth noting that $\lim_{\theta \to 0}\frac{2}{\theta}G(\frac{1}{2}{\theta})\gamma=\gamma$,
since we have shown in the proof of Proposition  \ref{prop:asympt} that $\lim_{x\to 0}\frac{G(x)}{x}=1$, and
that $\frac{2}{\theta}G(\frac{1}{2}{\theta})\sim \frac{2}{\mu(\gamma)^2}\log(\gamma \mu(\gamma)^2)$ as $\theta\to\infty$ by using 
(\ref{limit-Lambert}) and the  definition of $G$, thereby showing that the r.h.s. of (\ref{limit-general-E0-infinity-zero}) is a continuous function of $\lim_{\gamma\to\infty}\gamma\mu(\gamma)^2$ whenever this limit exists.
\end{remark}

\subsection{Numerical results}
\label{sec:brownian}

\begin{figure}[h]
\centering
\includegraphics[width=0.23\textwidth]{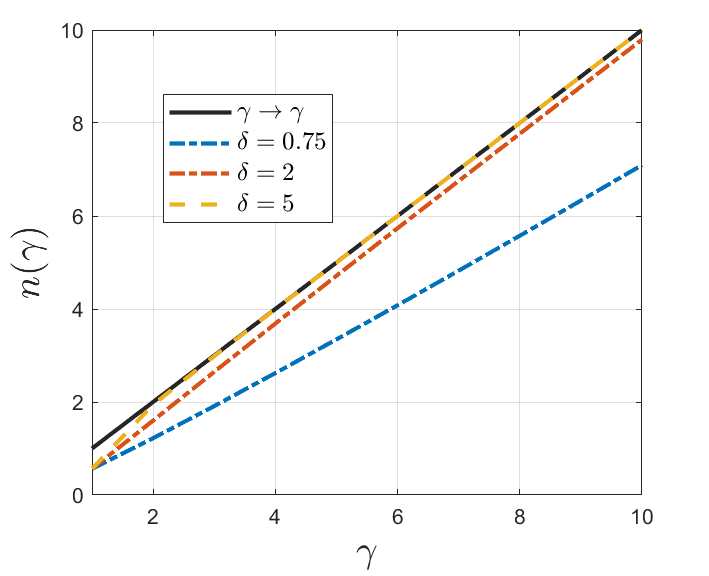}
\caption{The mapping $\gamma\to n(\gamma)$ when $\mu(\gamma)=\gamma^{-\delta}$ for $\delta\in\{0.75,2\}$ (dash-dotted lines), $\delta=5$ (dashed line),
and the identity
mapping $\gamma\to\gamma$ (solid line).}
\label{fig-n-gamma}
\end{figure}

Let $\mu(\gamma)=\gamma^{-\delta}$ with $\delta\geq 0$. We find from Proposition \ref{prop:asympt}
\begin{equation}
\label{limit-general-E0-infinity-zero-example}
n(\gamma) \sim
\left\{\begin{array}{ll}
2(1-2\delta) \log (\gamma)  \gamma^{2\delta}, &\mbox{if $0\leq \delta<\tfrac{1}{2}$,}\\
 2G(\frac{1}{2})\gamma,&\mbox{if $\delta=\tfrac{1}{2}$,}\\ 
 \gamma, &\mbox{if $\delta>\tfrac{1}{2}$,}
\end{array}
\right.
\end{equation}
as $\gamma\to\infty$, with $2G(\tfrac{1}{2})=0.563...$
The mapping $\gamma \to n(\gamma)$ (defined in (\ref{opt-solution}) and (\ref{E0}), with $h$ given in (\ref{h-star})) is displayed in Figure \ref{fig-n-gamma} for $\delta\in\{0.75,2,5\}$. We also plot
the identity mapping $\gamma\to \gamma$ so that one can observe the speed at which $n(\gamma)$ converges to $\gamma$, as predicted by
Proposition \ref{prop:asympt}. Not surpringly, 
the closer $\delta$ is to $\frac{1}{2}$ the longer it takes to converge to the asymptotic regime.
To better quantity this visual observation, we have reported in Table \ref{tab-M-gamma} the metric  $M(\gamma)=100\times |\frac{n(\gamma)-\gamma}{n(\gamma)}|$ for $\delta\in\{0.75,2,5\}$ and for  increasing values of $\gamma$. We observe that $n(\gamma)$ converges quickly to $\gamma$ for $\delta=2$ while
for $\delta=5$, $n(\gamma)$ matches $\gamma$ when $\gamma\geq 5$. Convergence is slow when $\delta=0.75$.

%

\begin{table}[h]
\begin{center}
\begin{tabular}{|| l | r | r | r | r | r | r  | r ||} \hline 
$\gamma=$ & $2$ & $5$ & $10$ & $10^2$ &$10^3$ &$10^4$ & $10^5$ \\ \hline
$\delta=0.75$  &63.7 &49.4 & 41.9&22.2 &12.2 & 6.78& 3.79\\  \hline
$\delta=2$   & 24.9&6.05 &2.12 &0.07 &0 &0 &0 \\  \hline
$\delta=5$& 2.97 & 0.04&0 &0 &0 &0  &0\\
 \hline
\end{tabular}
\end{center}
\caption{The metric $M(\gamma)$ for $\delta\in\{0.75,2,5\}$ and $\gamma\in\{2,5,10,10^2,\cdots,10^5\}$.}
\label{tab-M-gamma}
\end{table}

\begin{figure}[h]
	\centering
	\begin{tabular}{cc}
	\includegraphics[width=0.20\textwidth]{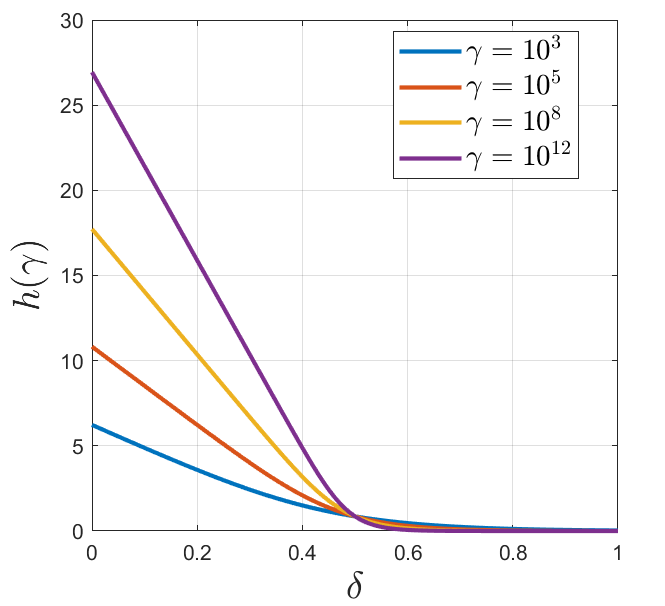} &
	\includegraphics[width=0.20\textwidth]{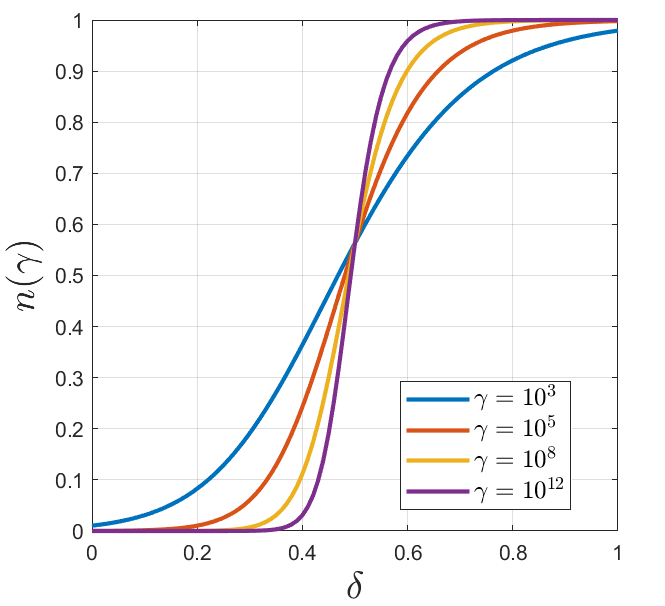}\\
		(a) & (b) \\
	\end{tabular}
	\caption{Phase transition analysis. As $\delta$ increases, (a) $h(\gamma)$ approaches zero and (b) the ratio $n(\gamma)/\gamma$ increases from 0 to 1, with the transition occurring at $\delta=0.5$.}
	\label{fig:phase}
\end{figure}

Figure~\ref{fig:phase}(a) shows how the optimal threshold $h(\gamma)$, given in (\ref{h-star}), 
varies with $\delta$ over the interval $[0, 1]$ for $ \gamma \in \{10^3, 10^5, 10^8, 10^{12}\}$. Similarly, Figure~\ref{fig:phase}(b) illustrates the behavior of $n(\gamma)/\gamma$ as a function of $\delta$, for the same values of $\gamma$.
We observe that a clear phase transition emerges as $\delta$ crosses 0.5, which corresponds to $\mu(\gamma) = 1/\sqrt{\gamma}$. This 
observation aligns with our theoretical findings.

\begin{figure}[h]
	\centering
	\includegraphics[width=0.20\textwidth]{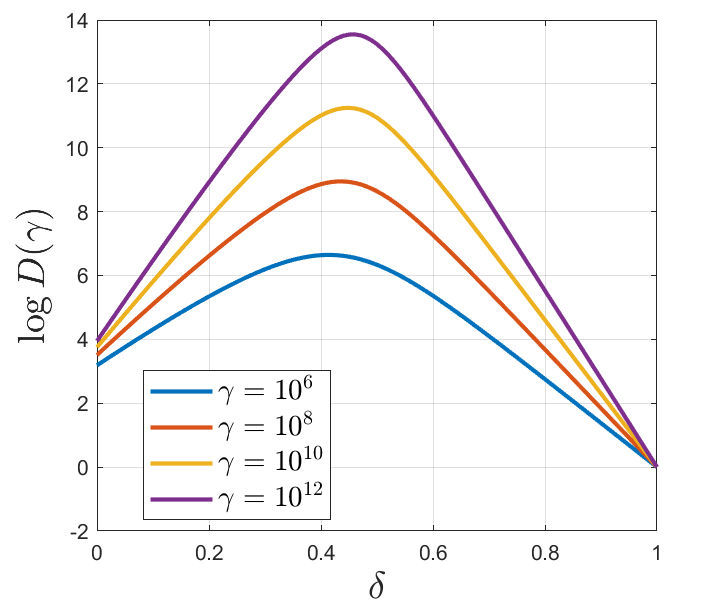}
	\caption{Total damage analysis.}
	\label{fig:damage}
\end{figure}

Now, we define the adversarial damage as $D(\gamma)=\mu(\gamma) n(\gamma)$. This quantity captures the cumulative impact an adversary can inflict before being detected. Following \eqref{limit-general-E0-infinity-zero-example}, 
\[
D(\gamma) \sim
\left\{
\begin{array}{ll}
2(1 - 2\delta) \log(\gamma) \gamma^{\delta}, & \text{if } 0 \leq \delta < \tfrac{1}{2}, \\
2G(\tfrac{1}{2}) \gamma^{1/2}, & \text{if } \delta = \tfrac{1}{2}, \\
\gamma^{1 - \delta}, & \text{if } \delta > \tfrac{1}{2},
\end{array}
\right.
\]
as $\gamma\to\infty$. From these expressions, $D(\gamma)$ is asymptotically maximized by $\delta = \tfrac{1}{2}$, resulting in a total damage of order $\Theta(\sqrt{\gamma})$. Figure \ref{fig:damage} displays the mapping $\delta \to \log D(\gamma)$ for $\gamma \in \{10^6,10^8,10^{10},10^{12}\}$.
The maximum damage occurs for $\delta \le \frac{1}{2}$; moreover, as $\gamma$ increases, the maximizer converges to $\delta = \frac{1}{2}$.

\section{Conclusion}
\label{sec:conclusion}
We introduced and analyzed a sequential detection problem when a covert adversary is present in a continuous-time setting where the post-change drift diminishes with the false alarm constraint parameter $\gamma$. By leveraging classical performance characterizations of the CuSum procedure, we rigorously established asymptotic expressions for the detection delay under vanishing drift. Our results identify the critical scaling regimes that govern covertness and show that adversarial damage is maximized when the drift decays as $\Theta(1/\sqrt{\gamma})$. Numerical simulations were included to illustrate and support the theoretical findings.



\appendix
\label{app}

{\it Proof of Lemma \ref{lem:solving-constraint}:}
Let  $\gamma>0$. Notice that $\xi(\gamma):=\frac{1}{2}\mu(\gamma)^2\not=0$ since, by assumption, $\mu(\gamma)>0$ as $\gamma>0$.
This shows that $h=0$ is not a solution of $e^{h}-h -1=\gamma \xi(\gamma)$.
Let us find a strictly positive solution $h$ to the equation 
\begin{equation}
\label{app:to-solve}
e^h-h -1 =  \gamma \xi(\gamma).
\end{equation}
 Eq.  (\ref{app:to-solve}) rewrites
\begin{equation}
\label{eq-in-y}
-(h+r)e^{-(h+r)}=-e^{-r},
\end{equation}
with $r:=1+\gamma \xi(\gamma)>1$. Since $-e^{-1}\leq -e^{-r}<0$, (\ref{eq-in-y}) has two solutions, $h_1=-r-W_0(-e^{-r})$ and $h_2=-r -W_{-1}(-e^{-r})$. 
From (\ref{branches-Lambert}), $h_1 <-\gamma \xi(\gamma)<0$, thereby implying  that the  solution we are looking for can only be $h=h_2$, namely,
\begin{equation}
\label{final-h}
h=-1 -\gamma \xi(\gamma)- W_{-1}(- e^{-1-\xi(\gamma)\gamma}).
\end{equation}
It remains to check that $h$ in (\ref{final-h}) is strictly positive for $\gamma>0$.

This will be true if one shows that the mapping $\varphi:x\to x- W_{-1}(- e^{x})$ is strictly positive for $x< -1$. We have $\varphi(-1)=0$ since
$W_{-1}(-e^{-1})=-1$. On the other hand,  by (\ref{derivative-W-1}),
\[
\varphi^\prime(x)=1+e^xW_{-1}^\prime(- e^{x})= \frac{1}{1+W_{-1}(-e^x)}, \quad \forall x<-1.
\]
But $W_{-1}(-e^x)<-1$ for $x<-1$ from (\ref{branches-Lambert}), which shows that $\varphi^\prime(x)<0$ for $x<-1$. Therefore, $x\to \varphi(x)$ is decreasing
 in $(-\infty,-1)$ and since $\varphi(-1)=0$, this proves that $\varphi(x)> 0$ for $x< -1$.
 Therefore, $h$ in (\ref{final-h}) is the unique strictly positive solution of the equation  $\frac{1}{2}\mu(\gamma)^2 (e^{h}-h -1)=\gamma$ when $\gamma>0$.

\section*{References}

\def\refname{}

\end{document}